\documentclass[aps,pra,reprint,superscriptaddress]{revtex4-2}

\usepackage{graphicx}
\usepackage{color}

\usepackage{xspace}

\newcommand{\MBNExplorer}
{\textsc{MBN Explorer}\xspace}

\newcommand{\MBNStudio}
{\textsc{MBN Studio}\xspace}

\begin{document}

\title{Modeling the effect of ion-induced shock waves and DNA breakage with the reactive CHARMM force field}

\author{Ida Friis}
\affiliation{Department of Physics, Chemistry and Pharmacy, University of Southern Denmark, Campusvej 55, 5230 Odense M, Denmark}

\author{Alexey V. Verkhovtsev}
\altaffiliation{On leave from Ioffe Institute, Polytekhnicheskaya 26, 194021 St. Petersburg, Russia}
\affiliation{MBN Research Center, Altenh\"oferallee 3, 60438 Frankfurt am Main, Germany}

\author{Ilia A. Solov'yov}
\altaffiliation{On leave from Ioffe Institute, Polytekhnicheskaya 26, 194021 St. Petersburg, Russia}
\affiliation{Department of Physics, Carl von Ossietzky Universit\"at Oldenburg, Carl-von-Ossietzky-Str. 9-11, 26129 Oldenburg, Germany}

\author{Andrey V. Solov'yov}
\altaffiliation{On leave from Ioffe Institute, Polytekhnicheskaya 26, 194021 St. Petersburg, Russia}
\affiliation{MBN Research Center, Altenh\"oferallee 3, 60438 Frankfurt am Main, Germany}

\begin{abstract}
Ion-induced DNA damage is an important effect underlying ion beam cancer therapy. This paper introduces the methodology of modeling DNA damage induced by a shock wave caused by a projectile ion. Specifically it is demonstrated how single- and double strand breaks in a DNA molecule could be described by the  reactive CHARMM (rCHARMM) force field implemented in the program \MBNExplorer. The entire work flow of performing the shock wave simulations, including obtaining the crucial simulation parameters, is described in seven steps. Two exemplary analyses are provided for a case study simulation serving to: (i) quantify the shock wave propagation, and (ii) describe the dynamics of formation of DNA breaks. The paper concludes by discussing the computational cost of the simulations and revealing the possible maximal computational time for different simulation setups.
\end{abstract}

\maketitle

\section{Introduction}

Classical molecular dynamics (MD) enables studying of structure and dynamics of
biological systems and inorganic materials, nowadays constituting of up to $10^6 - 10^8$ atoms, on
the nanosecond time scale \cite{Perilla_2015_CurrOpinStructBiol}.
In biological systems, which are in focus of this work, the interaction between constituent atoms in such systems is commonly described
through empirical molecular mechanics force fields,
for example, CHARMM \cite{MacKerell_1998_JPCB.102.3586, mackerell2000development}
or AMBER \cite{Cornell_1995_JACS.117.5179},
which are governed by the type of individual atoms and by the network of chemical bonds between them.
Being based on harmonic potentials, such force fields are well suited for modeling non-reactive
conformational changes in (bio)molecular systems, but cannot be used to simulate processes of breaking
and formation of chemical bonds which involve changes in atom connectivity.
Reactive force fields can overcome this limitation and enable simulation of chemical reactions in large systems
which cannot be simulated by \textit{ab initio} methods.
Well-known examples of reactive force fields are ReaxFF \cite{vanDuin_2001_reaxFF,senftle2016reaxff} and
reactive empirical bond-order (REBO) potentials \cite{Brenner_1990_PRB.42.9458, Brenner_2002_REBO}.
In both approaches,
the interatomic potential describes reactive events through a bond-order formalism
where bond order is empirically calculated from interatomic distances.

Recently, the so-called reactive CHARMM (rCHARMM) force field \cite{Sushko_2016_EPJD.70.12}
was introduced and implemented in the \MBNExplorer software package \cite{Solovyov_2012_MBNExplorer}.
Being an important extension of the standard CHARMM force field \cite{MacKerell_1998_JPCB.102.3586,mackerell2000development},
rCHARMM permits classical MD simulations of a large variety of molecular systems
experiencing chemical transformations whilst monitoring changes of their topology and
molecular composition \cite{MBNbook_Springer_2017, Sushko_2016_EPJD.70.217, Sushko_2016_EPJD.70.12}.
It takes into account additional parameters of the system, such as
bond dissociation energy, bond order and the valence of atoms.
The functional form of interatomic interactions is also adjusted to account for the finite
dissociation energy of the chemical bonds \cite{Sushko_2016_EPJD.70.12}.
The rCHARMM force field is particularly applicable for simulating irradiation- and collision-induced damage
phenomena where breaking of chemical bonds plays an essential role.
It was used previously to study
thermal splitting of water \cite{Sushko_2016_EPJD.70.12}, ion-induced water radiochemistry \cite{deVera_2018_EPJD.72.147},
collision-induced fusion and fragmentation of C$_{60}$ fullerenes \cite{Verkhovtsev_2017_EPJD.71.212},
irradiation-induced fragmentation of an organometallic compound W(CO)$_6$ \cite{deVera_2019_EPJD.73.215} as well as
the formation and growth of metal nanostructures in the process of focused electron beam induced deposition \cite{Sushko_2016_EPJD.70.217}.

This paper reports the first application of rCHARMM for modelling chemical transformations in
complex biological systems.
The number of systems and phenomena that can be studied using this approach is very large and
cannot be described in one paper.
Therefore we focus here on an exemplar case study and apply rCHARMM to simulate an important effect
arising when an ion projectile propagates through a biological medium, namely thermomechanical damage
of a DNA molecule by an ion-induced shock wave.
Understanding radiation effects produced by charged projectiles traversing biological media is of
utmost importance for radiotherapy with ion beams \cite{schardt2010heavy,solov2017nanoscale}
and space radiation protection \cite{Durante_2011_RMP.83.1245}.
Since modeling of chemical transformations in DNA (and other complex biomolecular systems) involves many different
technical and computational aspects, we present a detailed methodology for modeling such effects by means of rCHARMM.

The formation of ion-induced shock waves on the nanometer scale was predicted theoretically \cite{surdutovich2010shock}
and studied computationally in a series of subsequent works
\cite{surdutovich2013biodamage, devera2016molecular, deVera_2018_EPJD.72.147, Yakubovich_2012_NIMB.279.135, ES_AVS_2019_EPJD.73.241}.
Ions traversing a biological medium can deposit large amounts of energy per unit path length and most of this energy
is used to ionize water molecules and eject low-energy (below 50 eV) secondary electrons \cite{Surdutovich_2014_EPJD.68.353}.
Most of these electrons transfer their energy to electronic excitations of the medium in less than a nanometer
on a femtosecond time scale \cite{surdutovich2015transport}.
This time is much shorter than the typical time for energy dissipation via the electron-phonon coupling mechanism,
which occurs on a few picosecond timescale \cite{surdutovich2015transport}.
As a result, the medium in a nanometer-size cylinder around the ion track is strongly heated thus providing conditions
for a ``strong cylindrical explosion''.
This creates a discontinuity in the initial conditions for pressure that starts propagating in the radial direction
away from the ion path.
This propagation also features discontinuities in density and collective flow velocity and is referred to as a shock wave \cite{LL_FluidMechanics_vol6, Zeldovich_ShockWaves}.

Two possible effects of ion-induced shock waves in the radiation damage scenario have been suggested \cite{surdutovich2013biodamage}.
The first is related to direct DNA damage (such as the creation of DNA strand breaks) by thermomechanical stress \cite{Yakubovich_2011_AIP.1344.230, surdutovich2013biodamage, devera2016molecular, fraile2019first, bottlander2015effect}.
The other effect is the significant role of shock waves in transporting reactive species
(such as hydroxyl radicals and solvated electrons) due to radial collective
flows initiated by them \cite{surdutovich2015transport, deVera_2018_EPJD.72.147}.

Earlier MD studies \cite{Yakubovich_2011_AIP.1344.230, surdutovich2013biodamage, devera2016molecular}
of possible bond breakage in DNA by ion-induced shock waves were done using the standard CHARMM force field.
In these non-reactive simulations, the potential energy stored in a certain DNA bond was monitored in time
as its length varies around its equilibrium distance \cite{surdutovich2013biodamage, devera2016molecular}.
When the potential energy of the bond exceeded a given threshold
(which is about 3 to 6 eV for covalent bonds in the DNA backbone \cite{Range_2004_JACS.126.1654}),
the bond was considered as broken.

A deeper understanding of the shock wave induced DNA damage requires a more systematic approach
implying the use of reactive force fields.
Such consideration involves the following aspects.
First, a stable DNA structure should be constructed.
Second, the force field describing the rupture of covalent bonds in the DNA
(or at least in a specific part of it, e.g. in the backbone) must be developed.
Finally, the process of shock wave formation and propagation through a molecular medium should be simulated.
Details of the whole computational protocol and interlinks between its constituent parts are presented and discussed in this paper.
The methodology reported here represents an alternative approach to other methods
(such as ReaxFF \cite{senftle2016reaxff,bottlander2015effect, Abolfath_2011_JPCA.115.11045})
enabling simulations of bond breakage and formation within the MD framework.

The impact of a shock wave on DNA can be characterized through the probability of strand breaks formation,
which can be used to quantify the amount of biodamage induced by the shock wave mechanism.
This paper demonstrates how such analysis can be carried out with the rCHARMM force field.
The method is versatile and can be used for characterizing the material damage caused by the propagation
of different ions through various media.
It is also demonstrated how other important observables, such as the number of molecular fragments created by
the shock wave impact on the speed of shock wave propagation in the medium, can be derived from the simulations.
This analysis is important for understanding the radiation damage with ions on a quantitative level,
focusing on particular physical, chemical, and biological effects that bring about lethal damage to
cells exposed to ion beams \cite{Surdutovich_2014_EPJD.68.353, Verkhovtsev_2016_SciRep.6.27654, Verkhovtsev_2019_CNano.10.4}.
The methodology described in the present paper can be applied in further computational studies
considering irradiation of DNA with different ions and different orientations between the ion's path
and the DNA molecule.

\section{Methodology}

Computational modeling of an ion-induced shock wave hitting DNA in an aqueous environment is a multi-step process. The simulations rely in particular on the rCHARMM force field \cite{Sushko_2016_EPJD.70.12,Sushko_2016_EPJD.70.217}, and in this section, details are provided for every step that is required to successfully perform simulations.

\subsection{Construction of the system (step~1)}

\begin{figure*}[tb!]
\centering
\includegraphics[width=0.7\textwidth]{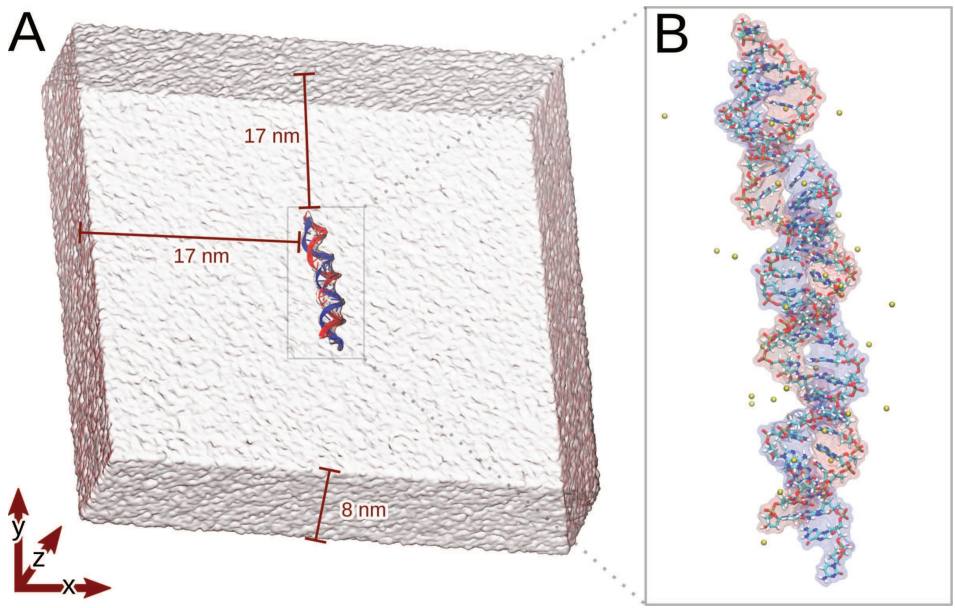}
\caption{ The molecular system used to simulate DNA damage by shock waves. Panel \textbf{A)} shows the equilibrated DNA double helix in a 34 nm$\times$ 34 nm$\times$ 8 nm water box. The dimensions of the water box along the $x$-, $y$-, and $z$-axis are indicated. In panel \textbf{B)} the DNA double helix strands are shown in greater detail with the neutralising sodium ions located near the DNA in yellow. }
\label{fig:eq system}
\end{figure*}

The study of a shock wave damaging DNA can be performed on different systems, but it is most natural here to consider a generic piece of DNA in a water box. Thus, the system is made from three joined DNA molecules (PDB-ID 309D \cite{qiu1997dna}) with the sequence (CGACGATCGT) which corresponds to the sequence (GCTGCTAGCA) in the complementary strand. For the sake of simplicity, the DNA molecule is placed in a pure water box representing the biological medium. The box extends 17 nm from the DNA to the edge of the water box in the $x$- and $y$-directions, as illustrated in Fig. \ref{fig:eq system}A. The dimensions of the water box are chosen here consistently with previous studies \cite{devera2016molecular}, where it was shown that a size of 17 nm is sufficient to simulate propagation of a shock wave generated by a carbon or iron ion on the timescale of approximately 10 ps. To neutralize the system, one sodium ion is placed for every phosphate group in the DNA to ensure a charge neutrality of the entire system. The total system size is 1,010,994 atoms and extends 8 nm in the \emph{z}-direction. After equilibration (described below) the sodium ions became distributed radially along the DNA, as expected, see in Fig. \ref{fig:eq system}B. The initial position of atoms could be generated through a number of computational tools. Among the most user-friendly ones VMD \cite{humphrey1996vmd} and \MBNStudio \cite{sushko2019modeling} can be named, which both have a number of plug-ins that can assist the construction of the DNA in solution.

\subsection{Classical molecular dynamics equilibration (step~2)}

\begin{figure*}[t!]
\centering
\includegraphics[width=0.6\textwidth]{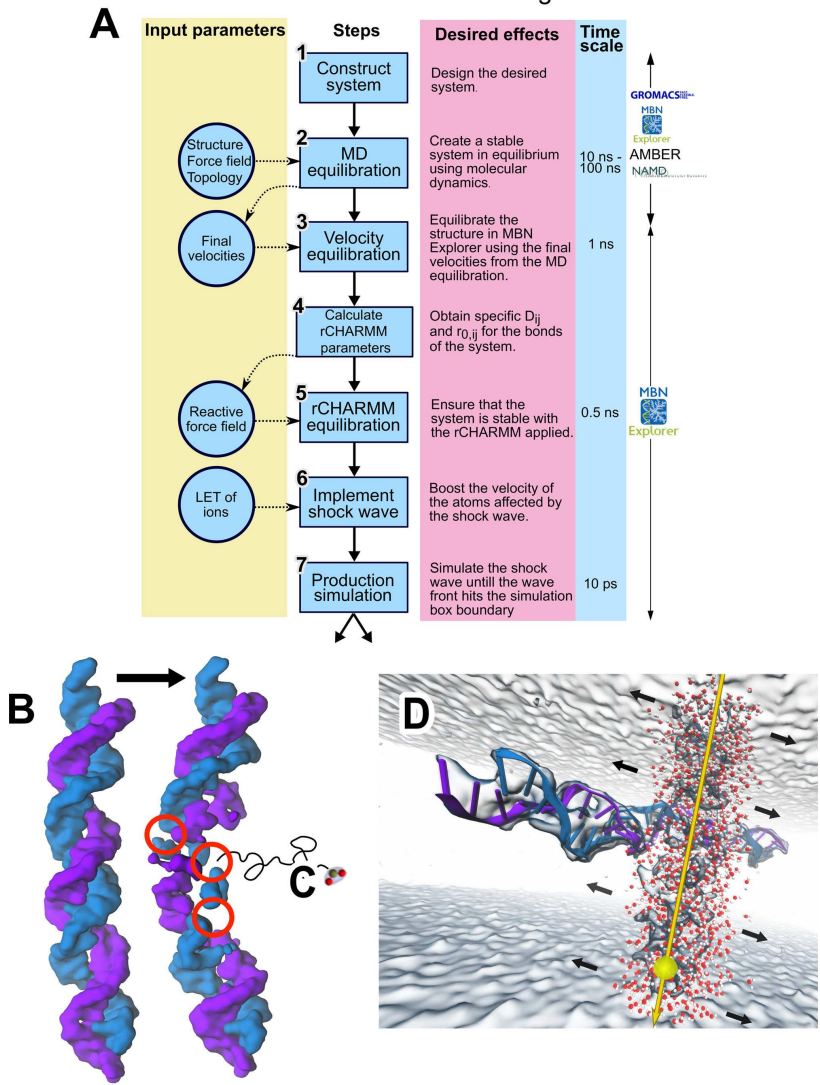}
\caption{The principal  workflow for simulations of ion-induced shock-waves propagating through a DNA molecule placed in a water box. The workflow relies on the reactive CHARMM force field (rCHARMM) \cite{Sushko_2016_EPJD.70.12,Sushko_2016_EPJD.70.217} procedure and includes seven steps. Panel \textbf{A} depicts the different steps needed before a simulation of a completely equilibrated system can be initiated (production simulation) and explains the purpose of every step. To the left side the input parameters are listed, while the right column depicts the characteristic simulation times. The bottom part illustrates several examples of results: Panel \textbf{B} shows three strand breaks being made to the DNA, panel \textbf{C} depicts ejection of a phosphate fragment from a DNA molecule and panel \textbf{D} illustrates the propagation of the ion-induced shock wave.}
\label{fig:flowchart}
\end{figure*}

Prior simulation of DNA bond breakage events, one has to ensure a stable, equilibrated configuration of the DNA fragment that will be subject to the shock wave. The process corresponds to step 2 in Fig. \ref{fig:flowchart}A. Equilibrium duration is dictated by the specific system of interest and could in principle be achieved with any MD software, e.g. NAMD \cite{phillips2005scalable}, AMBER \cite{case2016amber}, GROMACS \cite{van2005gromacs} or \MBNExplorer \cite{Solovyov_2012_MBNExplorer}. For larger systems and long equilibration times it may be advantageous to use NAMD, which for example provides significant speed up on GPU supercomputing clusters \cite{kindratenko2009gpu}. The standard CHARMM potential \cite{MacKerell_1998_JPCB.102.3586,mackerell2000development} should be used to model interactions within the DNA and with its water surroundings. The exemplary protocol used in the present study to obtain equilibrium is summarized in Table \ref{tab:equilibration}. In this paper, a small, flexible DNA piece solvated in a large water box is used, which makes the equilibration time relatively short, i. e. only 7 ns suffice. Note that the equilibration step where all constraints are lifted can potentially require hundreds of nanoseconds of simulation before the structure is stable, as for example seen when simulating cellular membranes, large protein structures or polymeric compounds \cite{husen2017mutations,barragan2016mechanism,beltukov2019modeling}.

\begin{table}[t!]
\centering
\begin{tabular}{c|c}
    \hline
    Constraints & Time  (ns)\\ \hline
    Not water and ions & 1.0 \\
    Backbone & 2.0\\
    None & 7.0 \\ \hline
\end{tabular}
\caption{An overview of the different equilibration procedures imposed on the DNA system, and applied in step 2 of Fig. \ref{fig:flowchart}A. In the first step everything but the water and sodium ions were constrained to allow relaxation of the water box. In the next step, the side chains of the DNA are equilibrated while the backbone atoms are constrained. For the final equilibration step, the entire system is released and allowed to equilibrate.}
\label{tab:equilibration}
\end{table}

Equilibration phase required periodic boundary conditions to be employed and an integration time step of 1 fs was used. The simulations were performed assuming the temperature of 310 K by utilizing the Langevin temperature control with a damping coefficient of 5 ps$^{-1}$ as well as Nose-Hover-Langevin piston pressure control \cite{feller1995constant} with a period of 200 fs and a decay time of 50 fs, keeping the average pressure of the system at 1 atm. Particle mesh Ewald summation \cite{darden1993particle} was employed for all electrostatic interactions and van der Waals forces were calculated with a cut-off distance of 12 \AA.

\subsection{Velocity equilibration (step~3)}

If the equilibration phase (Fig.~\ref{fig:flowchart}, step 2) was not performed using \MBNExplorer, but utilized e.g. NAMD, the equilibrated DNA structure should be transferred to \MBNExplorer \cite{Solovyov_2012_MBNExplorer}. This software has the unique feature that permits simulation of covalent bond rupture and all follow-up steps of the workflow can be accomplished by \MBNExplorer, see Fig.~\ref{fig:flowchart}A. In order to migrate the structure successfully, i.e. to avoid artifacts between the two programs, the final velocities for every atom need to be transferred from  NAMD, (AMBER or GROMACS) to \MBNExplorer as illustrated in step 3 of Fig.~\ref{fig:flowchart}A. It is important to note that turning the Langevin thermostat on at this point will induce a new velocity distribution for all atoms in the system that would follow a Boltzmann-distribution \cite{Solovyov_2012_MBNExplorer}, which would negate the act of transferring the final velocity from step 2 to step 3 (Fig.~\ref{fig:flowchart}A). Therefore, no thermostat should be used for the 1 ns simulation that aims to equilibrate the DNA-system after it has been transferred to \MBNExplorer. Naturally, step 3 of the computational protocol could be omitted if \MBNExplorer was used for the equilibration simulations in step 2.

{\subsection{Reactive CHARMM force field parameters (step~4)}
\label{sec:reactiveFF}}

The completely equilibtated DNA system obtained after step 3 (see Fig. \ref{fig:flowchart}A), is now ready for being parametrized using the rCHARMM force field. The following section details how the critical parameters are calculated. rCHARMM has the advantage that it can be implemented easily in \MBNExplorer with only addition of a single parameter namely the bond dissociation energy, as opposed to other reactive force fields, like ReaxFF in LAMMPS \cite{bottlander2015effect,senftle2016reaxff}.

The reactive force field represents a modification of the standard CHARMM force field \cite{MacKerell_1998_JPCB.102.3586},
that employs the harmonic approximation for the description of interatomic interactions thereby limiting its applicability
to small deformations of molecular systems.
Contrary to the standard CHARMM force field, its reactive modification goes beyond the harmonic approximation
for modeling covalent (bonded, angular and dihedral) interactions, and also accounts for a change of the molecular
topology, redistribution of atomic partial charges in the produced fragments as well as for atomic valences
of each fragment, therefore, enabling a more accurate description of the physics of molecular dissociation.

In the standard CHARMM force field \cite{MacKerell_1998_JPCB.102.3586} all covalent interactions are described
by a harmonic potential:
\begin{equation}
U^{\rm bond}_{\rm CHARMM}(r_{ij}) = k_{ij}^{\rm bond}  \left( r_{ij}-r_{0,ij}\right)^2 \ ,
\label{eq. harmonic}
\end{equation}
where $k_{ij}^{\rm bond}$ is the force constant of a bond, $r_{ij}$ is interatomic distance and $r_{0,ij}$ is the equilibrium bond length.
In the reactive force field \cite{Sushko_2016_EPJD.70.12}, the harmonic bond interaction is replaced with a Morse potential:
\begin{equation}
U^{\rm bond}(r_{ij}) = D_{ij}\left[ e^{-2\beta_{ij}(r_{ij}-r_{0,ij})}-2e^{-\beta_{ij}(r_{ij}-r_{0,ij})} \right] \mbox{ , }
\label{eq:Morse}
\end{equation}
where $\beta_{ij}=\sqrt{k_{ij}^{\rm bond}/D_{ij}}$ and $D_{ij}$ is the dissociation energy for the bond between atoms $i$ and $j$.
The parameter $\beta_{ij}$ determines the steepness of the potential.
The interactions are truncated at a user-defined cutoff distance that characterizes the distance beyond
which the bond is considered as broken and the molecular topology of the system changes.
The bond energy calculated by Eq.~(\ref{eq:Morse}) asymptotically approaches zero at large interatomic distances.
In order to ensure a continuous potential, switching functions are defined accordingly to gradually reduce the angular and dihedral interactions as the bond breaks \cite{Sushko_2016_EPJD.70.12}.

As described in the original paper \cite{Sushko_2016_EPJD.70.12}, once a bond starts to break, the associated angular and dihedral interactions fade away and eventually disappear entirely once the distance between atoms reaches a critical value. Once all the associated bonded, angular and dihedral interactions go to zero, they are automatically removed from the molecular topology of the system; the atoms that initially formed the broken bond are then considered as unbound, leading to formation of atoms with some dangling bonds. Upon the DNA strand break, the atoms with dangling bonds will interact through electrostatic and van der Waals interactions with their environment. It is important to mention that two atoms that were originally bound through a covalent bond do not experience any Coulomb or van der Waals interactions, as those interactions between neighboring atoms are automatically excluded in the CHARMM force field by definition. Once a bond is broken, the Coulomb and van der Waals interactions between the released atoms will appear. The impact of these interactions on fragment formations is, however, expected to be negligible: these interactions only appear once a bond is broken and then the atoms involved in the bonding would typically carry substantial kinetic energy, such that the involved fragments may propel away from each other. Thus, after some short time when the fragments relax, they are already so far from each other that a direct mending of the broken bond would be highly unlikely. If necessary, it is possible  to account for a more correct charge redistribution in the molecular fragments that arise upon bond rupture. Such a feature is possible in MBN Explorer, but it was not included in the present exemplary simulations, as it would not add much value to discussing the methodology of DNA strand breaking.

In the present paper, the use of rCHARMM force field for modeling direct thermomechanical damage of a DNA molecule due to the shock wave induced by a passing ion is presented. It is widely established that one of the key events of radiation-induced DNA damage concerns the formation of single- and double strand breaks (SSBs and DSBs) as well as more complex damages. Here our focus is primarily on the complex damage as it is considered to be of an irreparable type and is directly linked to cell survival probability. Therefore, the rCHARMM force field is used to describe covalent interactions in the DNA backbone while covalent interactions in other parts of the DNA molecule are modeled using the standard CHARMM force field. Below it is outlined how a limited set of parameters was obtained. The same approach should be used if more bonds are considered as reactive in the simulations.

\begin{figure}[tb!]
\centering
\includegraphics[width=0.45\textwidth]{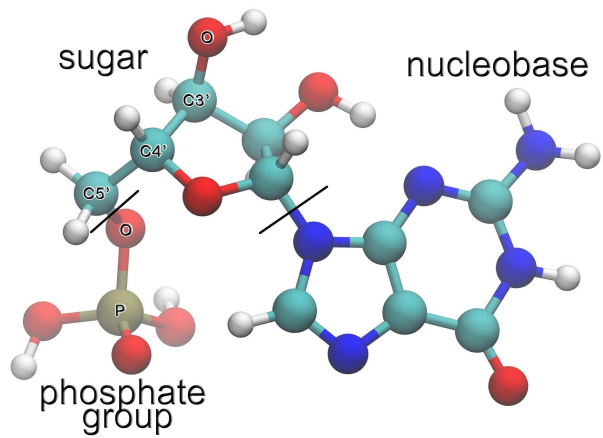}
\caption{Optimized geometry of guanosine monophosphate (GMP). Bonds between the labeled atoms have been considered
as reactive in the simulations. The black dashes split the molecule into three parts: the phosphate group, sugar part and the nucleobase.}
\label{fig:GMP_structure}
\end{figure}

Dissociation energy and cut-off distance parameters of rCHARMM for the specific bonds
were determined from density functional theory (DFT) calculations of a guanosine monophosphate (GMP, C$_{10}$H$_{14}$N$_5$O$_8$P) molecule,
whose structure is shown in Fig.~\ref{fig:GMP_structure}.
The calculations were performed using Gaussian 09 software \cite{Gaussian09} employing the B3LYP exchange-correlation
functional \cite{lee1988development,becke1988density} and the 6-31G(d,p) basis set for wavefunction expansion.
Geometry of a neutral GMP molecule was optimized first and then saved in the Z-matrix format to enable a fixed scan
over different covalent bonds.
For each bond a 50-step scan was performed in steps of 0.1~\AA, starting from the interatomic distance of about 0.6~\AA.
It was found that the covalent bonds of the DNA backbone, alongside with the glycosidic bond between the sugar ring
and the guanine nucleobase, have the lowest dissociation energies and thus likely have the higher probability to break.
The results for C3$^{\prime}$--O, C5$^{\prime}$--O, C4$^{\prime}$--C5$^{\prime}$ and P--O bonds of the DNA backbone
(see the labeled atoms in Fig.~\ref{fig:GMP_structure}) are listed in Table~\ref{tab: De parameters} and the corresponding parameters were used as input for computations in step 5, see Fig. \ref{fig:flowchart}A. It should be noted that a rupture of the C4$^{\prime}$--C3$^{\prime}$ bond will not cause a break in the DNA strand, as the bond is a part of a sugar ring, therefore it is not parametrized by the reactive CHARMM force field in this case.
Spring constants for the considered bonds are taken from the standard CHARMM force field for nucleic acids \cite{mackerell2000development}.

\begin{table}[t!]
    \centering
    \begin{tabular}{c|c|c| c |c}
         Type 1 & Type 2                              &   $D_{i,j}$ (kcal/mol)  & $r_{0,ij}$ (\AA) & $r_{\rm cutoff}$ (\AA) \\
         \hline
         C3$^{\prime}$ &O             & 160.22 & 1.415 & 3.60  \\
         C5$^{\prime}$ &O             & 160.22 & 1.415 & 3.60  \\
         C4$^{\prime}$&C5$^{\prime}$ & 146.71 & 1.518 & 3.90  \\
         P&O                         & 146.07 & 1.609 & 3.75  \\
    \end{tabular}
    \caption{Dissociation energy $D_{ij}$, equilibrium distance $r_{0,ij}$,
    and the cut-off distance $r_{\rm cutoff}$ for breakage/formation of covalent bonds in the sugar-phosphate backbone
    of guanosine monophosphate, calculated at the B3LYP/6-31G(d,p) level of theory. Cutoff distances are defined as the distances at which the bond energy is equal to $0.1 D_{ij}$.}
    \label{tab: De parameters}
\end{table}

\subsection{Reactive CHARMM force field equilibration (step~5)}

The parameters for the covalent bonds that are subject to break, and found in step 4, should now be included into the force field used for simulations. To ensure that the stability of the DNA system is not lost once the parameters for the DNA backbone have changed, a short simulation to relax the entire structure should be performed. It typically requires about half a nanosecond of simulation to allow for the bonds to settle into their new equilibrium distances and to ensure that no bonds will break unprovoked.

\subsection{Introducing the shock wave (step~6)}

The shock wave is induced by the ion travelling through the aqueous environment, where it loses its energy by electronic excitations and ionizations. A large amount of the ion's energy is deposited in a radius of 1 nm around the ion's track, generating the shock wave \cite{surdutovich2013dna,devera2019role,surdutovich2010shock}. To emulate this process, the shock wave is introduced to the final equilibrated structure of the system obtained from step 5 in the workflow (see Fig. \ref{fig:flowchart}A). It is assumed that energy deposited by the propagating ion is distributed uniformly among all atoms within the cylinder of 1 nm radius. Velocities of atoms inside this cylinder are increased by a factor $\alpha$ to match the energy deposited by the ion into the medium. This rescaling factor, $\alpha$, is found using the expression \cite{surdutovich2013biodamage}:

\begin{equation}
\sum_i^N \frac{1}{2} m_i \left( \alpha v_i \right)^2 = \frac{3 N k_{B} T}{2} + S l \ ,
\label{eq:velocity_scaling}
\end{equation}
where $S$ is the linear energy transfer (LET), that is the mean energy loss by an ion projectile per its unit path length,
$l$ is the length of the simulation box in the direction the ion is traveling and $N$ is the total number of atoms within the initial shock wave cylinder with length $l$. The second term on right hand side is the energy lost by the ion as it travels through the medium, and the first term is the kinetic energy of the cylinder at an equilibrium temperature, $T$.

LET depends on ion type and its initial kinetic energy.
The dependence of LET on ion's energy $E$ for different projectiles can be obtained
either from track-structure Monte Carlo simulations \cite{Friedland_2017_SciRep.7.45161, Rezaee_2018_RPOR.23.433}
or analytically (see Refs.~\cite{Surdutovich_2014_EPJD.68.353, Surdutovich_2019_CancerNano.10.6} and references therein).
The latter approach is based on the singly differential cross sections (SDCS) of ionization of water molecules with ions:
\begin{equation}
S(E) = - \frac{dE}{dx} = n \sum_i  \int\limits_0^{\infty} (W + I_i) \frac{d\sigma_i}{dW} dW \ ,
\end{equation}
where $n$ is the number density of water molecules in the medium and
$W$ is the kinetic energy of ejected electrons.
The sum on the right hand side is taken over all electron shells of the water molecule with $I_i$ being
the ionization potential and $d\sigma_i/dW$ the partial SDCS of each shell.

Having obtained the desired LET, the rescaling factor $\alpha$ can be found using Eq. (\ref{eq:velocity_scaling}) and used for rescaling the velocity of the water molecules within a radius of 1 nm of the ion track, effectively creating the “hot” cylinder.

\subsection{Production simulation (step~7)}

Once the system has been properly equilibrated and the shock wave correctly implemented, the time step in the production simulations needs to be tweaked. As the atoms within the hot cylinder are moving faster than in ordinary MD simulations, the integration time step needs to be shortened to prevent simulation stability issues. Here an integration timestep of 0.1 fs is usually a good starting choice, but even shorter time steps can be necessary to implement if the atoms are moving too fast. Once the simulation of the desired duration has been carried out, it can be proceeded with versatile analysis. Several examples of such an analysis are provided below to validate the overall protocol and methodology.

\section{Results}

The above described workflow has been employed to study shock wave propagation in a solvated DNA, as shown in Fig. \ref{fig:flowchart}A, which has undergone a production simulation for approximately 12-15 ps. As mentioned above, simulations with the reactive force field permit analysis of various characteristics, as for example illustrated in Fig. \ref{fig:flowchart}B,C and D. Below two examples of possible analysis of shock wave simulations are provided: \emph{(i)} a study of the shock wave propagation and \emph{(ii)} the temporal dependence of the total number of breaks emerging in the DNA strand. Note that the analysis of these examples is not given in all details, as they are used to provide a qualitative understanding of the  possibilities of the rCHARMM force field, since the main focus of this paper is placed on the description of the novel methodology.

\subsection{Shock wave propagation}

The shock wave propagates in the molecular system radially away from the ion track. The velocity of its wave front is the deciding factor for estimating the simulation time, as the system becomes non-physical once the boundary of the simulations box is hit. In this case the shock wave gets reflected and artificial interference arises.  Additionally, the medium density gradient of the wave front creates the force that potentially may lead to strand breaks in the DNA molecule. The strength of the shock wave depends on the LET considered. In this specific case study, an ion with an LET of 2890 eV/nm was used, corresponding to an argon ion at the Bragg peak.

\begin{figure*}[t!]
\centering
\includegraphics[width=0.85\textwidth]{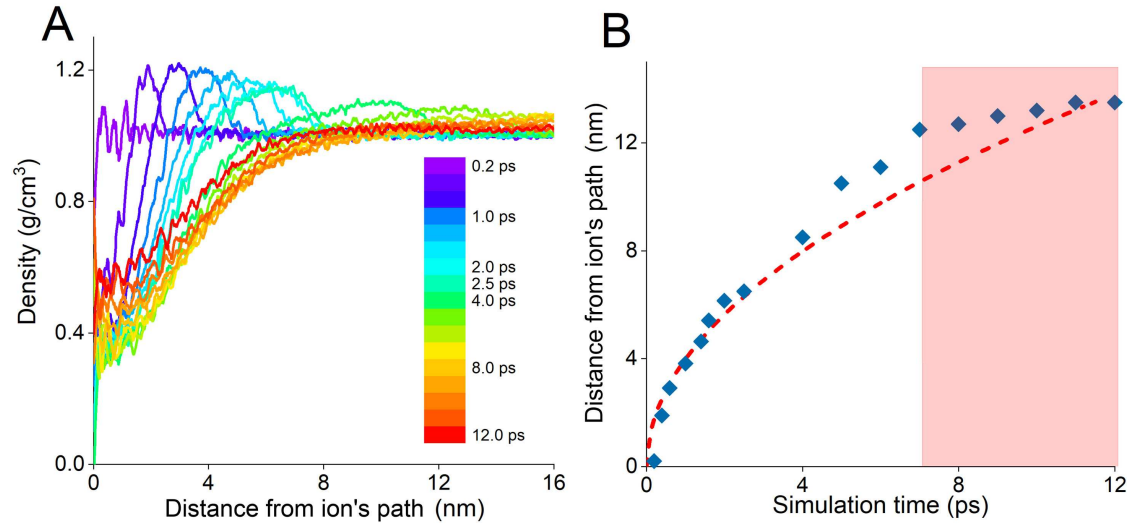}
\caption{Shock wave propagation through the solvated DNA, induced by an ion with LET of 2890 eV/nm, corresponding to argon. In panel \textbf{A)} the density of water in the radial direction from the ion beam center is shown at different instances, ranging from 0 ps to 12 ps after irradiation. In panel \textbf{B)} the location of the wave front is shown propagating as a function of simulation time and hitting the simulation box boundary at $\sim$8 ps. The wavefront propagates according to Eq. (\ref{eq. hydrodynamic model}), see dashed curve. The time domain where the simulation results match the analytical dependence corresponds to the situation before the wave front hits the boundary of the simulation box. The points within the red box are considered non-physical, as they are measured after the simulation box boundary has been hit. }
	\label{fig:wave_velocity}
\end{figure*}

The spread of the shock wave in the aqueous environment can be measured by observing the radial density of the molecules, as the shock wave evolves symmetrically around the ion track. The ion track was in this case placed directly through the geometrical center of the DNA strand. The results of this analysis are shown in Fig. \ref{fig:wave_velocity}A. The wave front moves toward the edge of the simulation box as time passes, and the wave profile becomes lower and broader, showing that the shock wave relaxes as time passes. This indicates that the impact of the shock wave lowers over time, and it can be seen in Fig. \ref{fig:wave_velocity}A that a DNA strand placed 2-4 nm away from the ion track is expected to receive the strongest  impact from the shock wave.

The radial position of the shock wave front changes with time according to the following relation:
\begin{equation}
R(t)= b \sqrt{t} \left( \frac{S}{\rho}\right)^{1/4} \ ,
\label{eq. hydrodynamic model}
\end{equation}
where $b$ = 0.86 for the specific cylindrical geometry, $S$ is the LET and $\rho$ is the density of unperturbed water.
\cite{surdutovich2010shock,devera2016molecular,surdutovich2013dna}.
When the wave front reaches the edge of the simulation box, it also reaches the physical limit of the simulation, where the natural propagation of the shock wave should be terminated. It can be seen in Fig. \ref{fig:wave_velocity}B, where it is apparent that the predicted location of the shock wave according to Eq. (\ref{eq. hydrodynamic model}) is discontinued when the front of the wave approaches the simulation box boundary. As such it is advisable not to continue simulations once the shock wave hits the boundary, if the results are meant to emulate realistic (bio)physical systems.

\subsection{Dynamics of strand break number}

\begin{figure}[t!]
\centering
\includegraphics[width=0.45\textwidth]{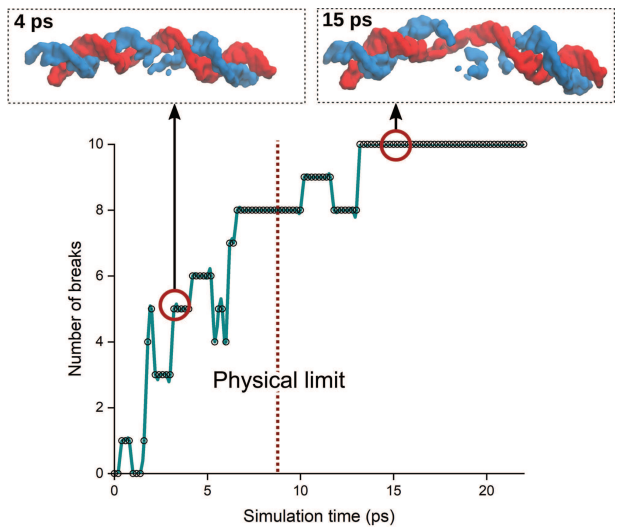}
\caption{The number of DNA strand breaks induced by a projectile ion with an LET of 2890 eV/nm, corresponding to an argon ion at the Bragg peak, computed as a function of time elapsed after the thermalisation phase. Two inserts in the top show the breakage of the DNA strands at two different time instances, one after a simulation time of 4 ps and another after a simulation time of 15 ps.}
\label{fig:breaks_vs_time}
\end{figure}

Even after the shock wave has passed the DNA, strand breaks can be formed as the difference in pressure continues to tuck on the DNA. This effect is illustrated in Fig. \ref{fig:breaks_vs_time}, where the number of breaks rises quickly within the first 6-7 ps of the simulation, where the wave front of the shock wave hits both DNA strands. This is the region where the shock wave has the most impact, as shown in Fig. \ref{fig:wave_velocity}. After this moment the  DNA damage rate becomes lower lasting until approximately 15 ps, hereafter the number of breaks remains steady. It can be seen that after the shock wave has hit the simulation box boundary it gets dispersed within approximately 8 ps, as seen in Fig. \ref{fig:wave_velocity}A, resulting in lowering the breakage rate.

Even if the number of breaks generally rises with time, some fluctuations of this quantity can be seen locally. This effect might be attributed to the broken bonds which can be rejoined if the atoms involved get close to each other after the initial bond breakage.

Other than strand ruptures, the shock wave causes deformation of the DNA, as seen in the inserts of Fig. \ref{fig:breaks_vs_time}, which might cause additional base pair breakage or misalignment thereby also damaging the structure of the DNA strand.

\section{Discussion}

The DNA damage caused by the ion-induced shock wave requires typically a couple of picoseconds to emerge \cite{devera2016molecular} and depends on the LET considered. The maximal simulation time, however, is limited by the distance from the ion track to the border of the water box. As the shock wave propagates in water, the wave front will travel towards the boundary of the simulation box. When it reaches the boundary where periodic boundary conditions are implemented, the shock wave will experience interference. This behavior is nonphysical and should be avoided, as it could introduce artifacts in the simulation results. Therefore if a long simulation is desired, a large surrounding water box is needed, however the choice of a larger system will make computations longer \cite{adcock2006molecular}. Below an estimate of the cost of shock wave simulations is provided.

\begin{figure*}[t!]
\centering
\includegraphics[width=0.85\textwidth]{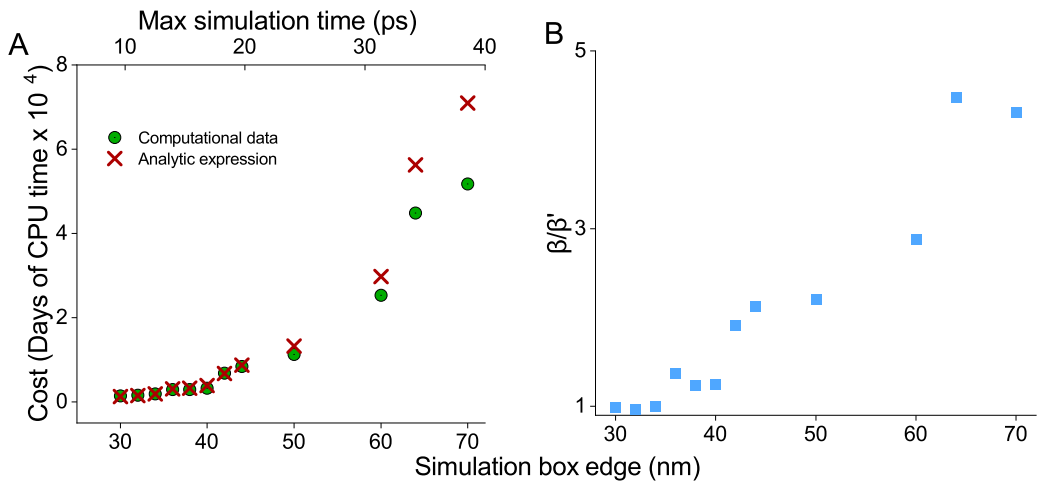}
\caption{A: The computational cost in CPU days. The green dots show the cost for the actual computed data points. The red crosses show the calculated cost according to Eq.~(\ref{eq. plotted cost}), where the ratio $\beta /\beta'$ is found from computations. The $\beta /\beta'$-values are shown to the right.
\label{fig:benchmarking}}
\end{figure*}

Assume a reference simulation box with the dimensions $L_x \times L_y \times L_z$. The total number of particles inside the simulation box for a given atomic density, $\rho$, can then be estimated as:
\begin{equation}
N = \rho L_x L_y L_z \ .
\label{eq. number atoms}
\end{equation}
Consider now that the simulation box is expanded by a distance $\Delta$ in the $x$- and $y$-directions. The corresponding number of atoms in the system, $N'$, then becomes:

\begin{equation}
N' = \rho (L_x+\Delta) (L_y+ \Delta) L_z \ .
\label{eq. N' xy}
\end{equation}
Assuming that the simulation box is square in the $x$- and $y$-direction, $L_x=L_y =L$, Eq. (\ref{eq. N' xy}) can be combined with Eq. (\ref{eq. number atoms}) and rewritten as:
\begin{equation}
N' = N\left( 1+ \frac{2 \Delta}{L} +\frac{\Delta ^2}{L^2}\right) \ .
\label{eq. N'}
\end{equation}
It is known that the shock wave front propagates as dictated by Eq. (\ref{eq. hydrodynamic model}), i. e. proportionally to the square root of the simulation time, see Fig. \ref{fig:wave_velocity}B. The maximal physically reasonable simulation time, $T_{\rm max}$, depends on the distance from the center of the shock wave to the simulation box boundary. Assuming that the ion passes through the box center, the reference value of the maximal simulation time becomes:
\begin{equation}
T_{\rm max}= \left( \frac{L}{2 a}  \right)^2 \ ,
\end{equation}
where $a = b \left( S  / \rho \right)^{1/4}$ is the proportionality factor between the radial distance of propagation for the wave front and $\sqrt{t}$, see Eq. (\ref{eq. hydrodynamic model}). For a simulation box increased by $\Delta$, the new maximal simulation time is found as:
\begin{equation}
T'_{\rm max}= \left( \frac{L+\Delta}{2 a}  \right)^2 = \left( 1+ \frac{ \Delta}{L}  \right)^2 T'_{\rm max} \ .
\label{eq. tmax}
\end{equation}
The computation time and the corresponding cost, $c$, of computing increase with the dimensions of the simulation box. The cost is measured in CPU days and depends on the number of picoseconds that can be obtained during one-day simulation on a single CPU core, here denoted as $\beta$:
\begin{equation}
c = \frac{T_{\rm max}}{\beta} n, \quad c' = \frac{T_{\rm max}}{\beta'} n' \ ,
\end{equation}
where $n$ is the number of CPU cores used in the reference calculations while $n'$ is the number of CPU cores in the case of a simulation with the increased dimensions. When increasing the system size, the computational cost of the simulations grows which typically means that the number of CPU cores may be increased by the user, such that:
\begin{equation}
N \rightarrow N', \quad n \rightarrow n' = n \frac{N'}{N}, \quad \beta \rightarrow \beta' \ .
\end{equation}
Using Eq.~(\ref{eq. N'}) and  Eq.~(\ref{eq. tmax}) and defining the increase in the simulation box size as $\Delta = L'-L$, the computational cost is found as:
\begin{equation}
\frac{T'_{\rm max}}{\beta'}n' \Big /  \frac{T_{\rm max}}{\beta}n
=
\frac{\beta}{\beta'} \left(\frac{L'}{L}\right) ^3 \ .
\label{eq. plotted cost}
\end{equation}
The system can thus be simulated for longer time scales if the surrounding water box is large enough. However, as follows from Eq.~(\ref{eq. plotted cost}), a large molecular system requires an increased  computational cost, as depicted in Fig.~\ref{fig:benchmarking}A.

The fraction $\beta/\beta'$ is shown in Fig.~\ref{fig:benchmarking}B; here $\beta$ corresponds to a reference system size of $L = 34$~nm, used as an example throughout this paper. In the ideal scenario, $\beta/\beta'$ should be close to 1, which, however, is not the case and noticeable derivations are seen. A comparison of the calculated cost of computation with the value taken from the actual benchmark simulations are shown in Fig.~\ref{fig:benchmarking}A.

The short simulation times are also governed by the rapid motion of the water molecules affected by the ion. When the atomic velocities are high, the integration step needs to be low in order to prevent error, making the computational and financial cost at least ten to twenty times higher for the shock wave simulations. Indeed, conventional MD methods  use an integration timestep of 2 fs \cite{sjulstok2018molecular,kimo2018atomistic,friis2017}, while shock wave simulation requires an integration time step of at most 0.1 fs.

The short simulation times do not, however, prevent the presented methodology in conveying significant results. It can be noted that within the first $\sim$10 ps, the shock wave has traveled the entire waterbox and DNA breaks have formed by a projectile argon ion. It can be observed that in this considered time frame, the DNA strand deforms and both SSBs and DSBs emerge. It can be noted that the relatively short simulations performed with the rCHARMM force field were successfully used in several other systems than the one presented in this paper: In a study of water splitting using the reactive force field, significant results were obtained in less than 100 ps \cite{Sushko_2016_EPJD.70.12}, and fullerene-fullerene collisions were simulated to an end within 10 ps \cite{Verkhovtsev_2017_EPJD.71.212}. Thus significant effects involving bond breakage and formation can be observed even on small time scales as the one used in the present work. However, some effects cannot be modeled at such short time intervals. For example the DNA molecule and its aqueous surroundings can not possibly reach a thermal equilibrium after the ion has passed by in just a few picoseconds. Further longer-times effects could involve processes after the settling of the shock wave, which require significantly longer times and, among other processes, can lead to rejoining of some covalent bonds.

\section*{Acknowledgments}
The authors are grateful for financial support from the Lundbeck Foundation, the Danish Council for Independent Research, the Volkswagen Stiftung (Lichtenberg professorship to IAS), and the DFG (Grant No. GRK1885 and No. 415716638), as well as from the European Union's Horizon 2020 research and innovation programme (H2020-MSCA-IF-2017 'Radio-NP', grant agreement No. 794733). The authors are also grateful to the DeiC National HPC Center (SDU) as well as DFG and the Ministry of Science and Culture of the State of Lower Saxony for providing the computational resources necessary for the calculations.

\bibliography{bibliography}

\end{document}